\documentclass[lettersize,journal]{IEEEtran}

\usepackage{amsmath,amsfonts}
\usepackage{array}
\usepackage[caption=false,font=normalsize,labelfont=sf,textfont=sf]{subfig}
\usepackage{textcomp}
\usepackage{stfloats}
\usepackage{url}
\usepackage{verbatim}
\usepackage{graphicx}
\usepackage{cite}
\usepackage{setspace}       % 1.5 spacing
\usepackage{multicol}
\usepackage{subfig}
 \usepackage{multirow}
\usepackage{color}
\usepackage{graphicx}
\usepackage{tikz}
\usepackage[export]{adjustbox}
\usepackage{enumitem}
\usepackage{comment}
\usepackage{comment}
\usepackage{float}
\usepackage{hyperref}
\usepackage{svg}
\usepackage{graphicx}
\usepackage{subfig}
\usepackage[linesnumbered,ruled,vlined,noend]{algorithm2e}
\usepackage[noend]{algcompatible}

\usepackage{mathabx}
\usepackage{color}
\usepackage{siunitx}
\usepackage{titlesec}
\usepackage{makecell} % Include this in your preamble
\usepackage{subfig}

\hypersetup{
    colorlinks,
    citecolor=black,
    filecolor=black,
    linkcolor=black,
    urlcolor=black,
    linktoc=all
}

\begin{document}

\title{Adaptive K-PackCache: Cost-Centric Data Caching in Cloud}

\author{\IEEEauthorblockN{Suvarthi Sarkar, Aadarshraj Sah, Poddutoori Sweeya Reddy, Aryabartta Sahu ~\IEEEmembership{IEEE Senior Member}}

\IEEEauthorblockA{\textit{IIT Guwahati, India.} Emails: s.sarkar@iitg.ac.in, \{s.aadarshraj,r.poddutoori\}@alumni.iitg.ac.in, asahu@iitg.ac.in}

\thanks{Aadarshraj Sah and Poddutoori Sweeya Reddy contributed equally to this work.}% <-this % stops a space
\thanks{Manuscript received April 19, 2021; revised August 16, 2021.}
}

% The paper headers
\markboth{Journal of \LaTeX\ Class Files,~Vol.~14, No.~8, August~2021}%
{Shell \MakeLowercase{\textit{et al.}}: A Sample Article Using IEEEtran.cls for IEEE Journals}

\IEEEpubid{0000--0000/00\$00.00~\copyright~2021 IEEE}
% Remember, if you use this you must call \IEEEpubidadjcol in the second
% column for its text to clear the IEEEpubid mark.

\maketitle

\begin{abstract}
Recent advances in data analytics have enabled the accurate prediction of user access patterns, giving rise to the idea of \textit{packed caching}—delivering multiple co-accessed data items together as a bundle. This improves caching efficiency, as accessing one item often implies the need for others. Prior work has explored only 2-item (pairwise) packing. In this paper, we extend the concept to general $K$-packing, allowing variable-size bundles for improved flexibility and performance. We formulate the K-PackCache problem from a content delivery network (CDN) operator’s perspective, aiming to minimize total cost comprising two components: (i) transfer cost modeled
as a base cost plus a linearly increasing term with the
number of items packed, and (ii) memory rental cost for caching, which depends on how long and how much is stored. Overpacking increases cost due to low utility; underpacking leads to missed sharing opportunities. We propose an online algorithm, \textit{Adaptive K-PackCache (AKPC)}, which dynamically forms, merges, and splits data cliques based on user access patterns and content correlation. Our approach supports batch requests, enables approximate clique merging, and offers a formal competitive guarantee. Through extensive evaluation on the Netflix and Spotify datasets, \textit{AKPC} reduces total cost by up to 63\% and 55\% over online baselines, respectively, and achieves performance within 15\% and 13\% of the optimal. This demonstrates its scalability and effectiveness for real-world caching systems.
\end{abstract}


%\setstretch{1.2}

%%%%%%%%%%%%%%%%%%%%%%%%%%%%%%%%%%%%%%%%%%%%%%%%%%%%%%%%%%%%%%%%%%%%%%%%%%%%%%%%
\section{Introduction}

\IEEEPARstart{T}{he} emergence of edge cloud systems has transformed how data is stored and accessed. By bringing data closer to users, these systems reduce latency and improve responsiveness. In edge environments, data is distributed across multiple nodes, cutting down the distance it needs to travel and enhancing access speed \cite{Roy23}. A key technique that supports this efficiency is data caching, which stores frequently accessed data near the point of use. Caching plays a vital role in improving performance in edge computing systems. Recently, a new caching paradigm called ``\textit{PackCache}'' has gained attention. This approach looks at data access patterns and identifies items that are often used together. Instead of handling each item separately, it bundles such items and sends them together. This reduces communication cost and makes caching more efficient at the edge \cite{Wu23}.

Nowadays, short videos on social media, such as reels and shorts, are gaining popularity. People often watch a series of these videos, typically scrolling through multiple similar ones in a single session \cite{Chen22}. A similar example is found in brief news applications like DailyHunt and Inshorts, where accessing a news article often leads to viewing related content, such as corresponding pictures and video clips, shortly after \cite{Huang19}. These scenarios differ from traditional recommendation systems, where users are presented with similar items based on previous user interactions. In these cases, the data is bundled together, as users frequently access related content within a very short interval. The concept of \textit{PackCache} is inspired by two observations: (a) recent data studies show that over 93\% of human behavior, including data access patterns in cloud services, is predictable in both spatial and temporal domains \cite{Song10}; and (b) cloud-based caching generally does not have strict limits on cache capacity. Instead, the cache size can be virtually unlimited, as long as users can afford it according to the cost model \cite{Wang17}.

Packed data caching is a sophisticated approach aimed at efficiently storing frequently co-utilized data items in cloud environments. It revolves around recognizing patterns of co-utilization among various data elements, enabling the development of effective offline and online caching approaches. 

In the existing literature, some research has been conducted using an offline approach on predicted data, employing greedy algorithms and dynamic programming techniques to pack data \cite{Huang19}. However, these methods can be unrealistic due to variations between real-time data and predicted data. Wu et al. \cite{Wu23} extended this work into the online domain using FP-tree structures, but their method was limited to packing at most two data points, hence we refer their work as 2-packing. Our methodology builds upon this prior research by extending the concept of 2-packing to accommodate the packing of $K$ data points. This extension allows for a more generalized and flexible approach to data caching, enhancing its adaptability and effectiveness in dynamic environments. By constructing a co-utilization graph and leveraging $K$-cliques, our approach optimizes data caching, and uses the cliques to address challenges related to data transfer and data access in cloud systems.

The key contributions of our work are summarized below:

\begin{itemize}
    \item \textit{Generalized $K$-Packing for Caching: } We extend traditional 2-packing strategies to a more flexible $K$-packing framework, allowing the grouping of arbitrarily many co-accessed data items into cliques of size up to $\omega$. Unlike prior works that process single requests in isolation, our approach supports batch-based request handling, which more accurately reflects practical traffic patterns in modern edge and cloud systems. This enables improved packing opportunities and better utilization of caching resources.

    \item \textit{Competitive Performance Guarantee: } We formally prove that our proposed \textit{Adaptive K-PackCache (AKPC)} algorithm is $\frac{2 + (\omega - 1) \cdot \alpha \cdot \mathcal{S}}{1 + (\mathcal{S} - 1) \cdot \alpha}$-competitive with respect to the offline optimal solution, where $\omega$ is the maximum clique size, $\alpha$ is the discount factor of transfer cost for packed data items, and $\mathcal{S}$ denotes the number of data items in a request that are not cached locally. Furthermore, we show that no deterministic online algorithm can achieve a better competitive ratio, establishing a tight bound.

    \item \textit{Approximate $K$-Clique Formation with Splitting and Merging: } To enhance packing efficiency and reduce fragmentation, we introduce two key modules: (a) \textit{Clique Splitting} — cliques larger than $\omega$ are recursively partitioned to adhere to the maximum clique size constraint. These mechanisms significantly improve the quality of packed groups while maintaining computational tractability; and (b) \textit{Approximate Clique Merging} — cliques of size $\omega$ are formed even when the induced subgraph has only a fraction $\gamma$ of the possible $\binom{\omega}{2}$ edges

    \item \textit{Empirical Gains and Real-World Applicability: } Our approach is validated on two real-world traces—Netflix and Spotify—and shows strong practical value. Compared to the best-known online state-of-the-art, \textit{AKPC} reduces total cost by up to 63\% and 55\% on Netflix and Spotify, respectively. While being an online method, it achieves near-optimal behavior—only 15\% and 13\% worse than the offline optimal baseline for Netflix and Spotify, respectively—demonstrating its robustness and scalability in real-world settings.
\end{itemize}

\section{Review of Prior Works}

The data caching problem is typically studied under two paradigms: \textit{offline} and \textit{online}. In offline settings, the entire request sequence is known in advance, whereas online approaches must operate without future knowledge. Several works have investigated efficient caching strategies and network-aware control mechanisms. Cohen \textit{et al.} ~\cite{Itamar25} addressed the problem of adaptive cache selection and content advertisement by introducing SALSA2, a learning-based framework that jointly estimates false-indication probabilities and adjusts the frequency and size of indicator broadcasts. Building upon ideas of cache partitioning and modularity, Park \textit{et al.} ~\cite{Park24} proposes Divide and Cache, a control plane framework for private 5G networks. By splitting network functions between on-premise and cloud locations based on dependency and frequency analysis, and introduced a Network Function Profile Cache Function (NFPCF).

In a broader context, the work by D'Alconzo \textit{et al.} ~\cite{DAlconzo19} surveys big data applications in network traffic monitoring, highlighting challenges in scalability, anomaly detection, and real-time processing — all of which are relevant to cache and control management in large-scale systems.

The security dimension of data caching and offloading is explored in Gu \textit{et al.} ~\cite{Gu23}, which introduces a dual attribute-based auditing mechanism for fog-assisted dynamic storage systems. Similarly, Zheng \textit{et al.} ~\cite{Zheng25} presents DIADD, a secure and efficient duplication and data integrity auditing scheme for cloud storage with support for data dynamics, addressing both scalability and security.

At a more system-level optimization front, Tajiri \textit{et al.} ~\cite{Tajiri25} studies federated learning over edge-cloud networks and proposes bandwidth-efficient data/model transfer strategies. Complementarily, Tian \textit{et al.} ~\cite{Tian23} proposes a federated deep reinforcement learning-based cooperative caching framework that jointly optimizes hit ratio and access delay across edge nodes.

A recent advancement by Huang \textit{et al.}~\cite{Huang19} introduced \textit{DP\_Greedy} which combined dynamic programming and greedy strategies to introduce packing-based caching in cloud systems. However, their approach is confined to offline settings and only considers pairwise (2-packing) grouping. Most prior methods either lack a packing component, limit packing to two data items, or are restricted to offline scenarios with complete request knowledge.

Wu \textit{et al.}~\cite{Wu23} introduced \textit{PackCache}, the first online packing-based caching algorithm. It uses FP-Trees to identify frequently co-accessed data items and employs anticipatory caching to manage cache contents in an online setting. While this represents a significant step toward dynamic packing, it is limited to fixed-size (pairwise) packing and single-request handling.

We extend online packing by generalizing pairwise packing to $K$-packing, allowing flexible grouping of co-accessed items for better cache efficiency. Unlike prior work focused on single-item requests, we model realistic batch arrivals. This advances existing literature by bridging theoretical models with practical, scalable caching solutions for modern cloud and edge environments.

\nocite{nc1}
\section{System Model and Problem Formulation}
\begin{figure}[tb!]
\centering
\includegraphics[scale=.8]{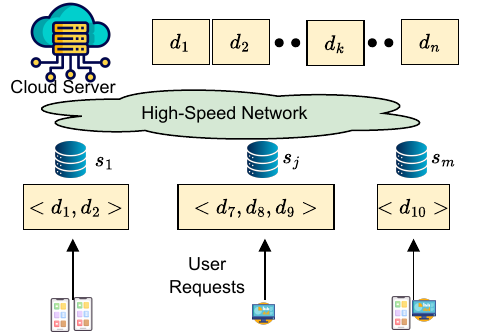}
\caption{\footnotesize System Architecture of the proposed model}
\label{fig:schema}
\end{figure}

We consider a Content Delivery Network (CDN) architecture, as illustrated in \autoref{fig:schema}, designed to deliver data efficiently to end users. The architecture comprises a central cloud server and multiple edge storage servers (ESSs), all interconnected via a high-speed network. Users send data requests to their designated ESS, which serves the request either locally or by retrieving the required data from other ESSs or the cloud server.

\subsection{Storage Model}

The CDN rents computational and storage resources from cloud infrastructure providers, incurring costs proportional to the resources consumed. Specifically, it rents a central cloud server and a set of edge storage servers denoted by $\mathcal{S} = {s_1, s_2, \ldots, s_j, \ldots, s_m}$, where $s_j$ represents the $j^{th}$ edge storage server and $m$ is the total number of ESSs in the system (so $|\mathcal{S}|=m$). Each ESS is assumed to have virtually unlimited storage capacity, constrained only by the rental allocation. All ESSs are fully interconnected via a high-speed internal network.

The set of all data items is represented by $\mathcal{U} = {d_1, d_2, \ldots, d_k, \ldots, d_n}$, where $d_k$ denotes the $k^{th}$ data item and $n$ is the total number of items (so $|\mathcal{U}|=n$). All data items are of uniform size, and the system provides read-only access to the users to ensure consistency and reduce complexity.

The cloud server maintains a persistent, unpacked copy of every data item in $\mathcal{U}$, ensuring durability and availability. In contrast, edge storage servers act as fast-access caches and store data items in either packed or unpacked formats based on recent access patterns. For example, suppose that data items $\langle d_1, d_2 \rangle$ and $\langle d_7, d_8, d_9 \rangle$ are frequently co-utilized. These items are then packed together and stored as a single unit.

If a user associated with $s_j$ requests $d_1$, the packed bundle $\langle d_1, d_2 \rangle$ is fetched (if not already cached) and cached in $s_j$, and then served to the user. Similarly, if any one of the data items among ${d_7, d_8, d_9}$ is requested by a user of $s_j$, the full packed group denoted as $\langle d_7, d_8, d_9 \rangle$ is fetched and cached in $s_j$. Then it is served to the user. Conversely, if a user under $s_j$ requests a non-co-utilized item such as $d_{10}$, which is not part of any packed group, only $\langle d_{10} \rangle$ is fetched and cached individually in $s_j$.

This hybrid approach enables the CDN to optimize both cost and latency by strategically managing packed data placement in the ESSs, while using the cloud server for guaranteed availability and fallback access.

\subsection{Data Request Model}

Users access data from the CDN by sending requests to the edge storage server (ESS) located closest to them. Each data request $r_i$ is modeled as a tuple $\langle D_i, s_j, t_i \rangle$, where $D_i$ denotes the set of data items requested, $s_j$ represents the $j^{th}$ ESS at which the request is made, and $t_i$ denotes the time instance at which the request is generated by the user.

The request set $D_i$ can contain either a single data item or multiple data items, with the number of requested items ranging from $1$ to a predefined upper bound $d_{\max}$. If a request includes the maximum number of items, the set can be expressed as $D_i = \{d_i^1, d_i^2, \ldots, d_i^{d_{\max}}\}$, where each data items $\in \mathcal{U}$, the universal set of all data items.

The temporal dimension of requests is bounded by a maximum time horizon $T_{\max}$, such that $t_i \in [1, T_{\max}]$. Each data item is associated with a fixed expiration interval $\Delta t$, meaning that once it is accessed and cached on an edge storage server, it remains available for the next $\Delta t$ time units unless re-accessed. If the item is re-accessed within this window, its expiry is extended by another $\Delta t$ from the latest access time. We also assume that each server is capable of handling multiple incoming requests concurrently.

\subsection{Cost Model}

We adopt the cost model proposed in \cite{Huang19,Wu23}, which captures two primary types of expenses incurred by the CDN: (a) Caching Cost denoted by $C_P$. It is the cost paid by the CDN to ESSs for renting storage space. (b) Transfer Cost denoted by $C_T$, It is the cost incurred for transferring data items across ESSs or between the cloud and ESSs, paid to the network service provider. 

\begin{figure}
    \centering
    \includegraphics[scale=0.9]{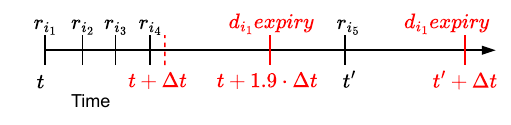}
    \caption{\footnotesize Visualization of time-window-based caching where all the requests want access of data item $d_{i_1}$.}
    \label{fig:cache_example}
\end{figure}

Consider the scenario illustrated in \autoref{fig:cache_example}, where a request $r_{i_1}$ for data item $d_{i_1}$ arrives at time $t$ at edge storage server (ESS) $s_j$. As a result, $d_{i_1}$ is cached at $s_j$ with an initial expiry time of $t + \Delta t$, according to the default caching policy.

Subsequently, additional requests $r_{i_2}$, $r_{i_3}$, and $r_{i_4}$ for the same data item $d_1$ arrive within the caching window $[t, t + \Delta t]$. Specifically, $r_{i_2}$ and $r_{i_3}$ arrive before $t + 0.9 \Delta t$, and $r_{i_4}$ arrives exactly at $t + 0.9 \Delta t$. Each of these requests extends the expiry of $d_1$ by an additional $\Delta t$ from their respective access times. Thus, after $r_{i_4}$, the expiry is updated to $t + 1.9 \Delta t$. All intermediate requests (e.g., $r_{i_2}$, $r_{i_3}$) benefit from zero caching cost, as $d_1$ remains cached throughout. However, the caching cost is attributed to the last request that extends the cache duration—in this case, $r_{i_4}$. Hence, the total caching cost for maintaining $d_1$ until $t + 1.9 \Delta t$ is charged.

Later, a new request $r_{i_5}$ for $d_1$ arrives at time $t' > t + 1.9 \Delta t$, after the extended cache has expired. Since $d_1$ is no longer present in the cache, the data item needs to be fetched and cached till expiry ($t'+\Delta t$).

Let $\mu$ denote the caching cost per data item per unit time. For a request $r_i$, which involves a data set $D_i$, the caching cost incurred during a time interval of duration $\Delta t$ is defined as:

{\footnotesize \begin{equation}
C_P^{r_i} = |D_i| \cdot \mu \cdot \Delta t,
\end{equation}}

where $|D_i|$ represents the number of data items associated with request $r_i$. For singleton requests (i.e., $|D_i| = 1$), this reduces to a simple per-item cost of $\mu \cdot \Delta t$. The total caching cost across all requests is given denoted by $C_P$ and is defined as:

{\footnotesize \begin{equation}
C_P = \sum_{r_i \in R} C_P^{r_i}.
\end{equation}}

We similarly define the transfer cost using a base transfer rate $\lambda$, which denotes the cost of transferring a single data item per unit time. For requests where data items are transmitted individually, the cost is linear in the number of items, i.e., $|D_i| \cdot \lambda$.

To capture the cost savings from co-accessed items being transmitted together, we introduce a discount factor $\alpha \in [0, 1]$, representing the degree of overlap or compression in packed transmission. Under this model, the total transfer cost for a packed request $r_i$ involving $|D_i|$ items is given by:

{\footnotesize \begin{equation} C_T^{r_i} = \begin{cases} (1 + (|D_i| - 1) \cdot \alpha) \cdot \lambda & \text{(Packed)} \\ |D_i| \cdot \lambda & \text{(Unpacked)} \end{cases} \label{eq:transfer_cost} \end{equation}}

Note that for $\alpha < 1$, packed transmission is always cheaper than transmitting items independently. The aggregate transfer cost across all requests is denoted by $C_T$ and is defined as:

{\footnotesize 
\begin{equation}
C_T = \sum_{r_i \in R} C_T^{r_i}.
\end{equation}}

This formulation allows us to jointly evaluate both caching and transfer costs under various packing strategies and co-access patterns.

The discount factor $\alpha$ allows fine-grained control over the trade-off between packing efficiency and transmission cost. A lower value of $\alpha$ encourages packing, thereby reducing $C_T$ and improving overall system efficiency. A detailed comparison of caching and transfer costs for varying request sizes ($|D_i|$) is provided in \autoref{data_table}.

\begin{table}[tb!]
\tabcolsep1.7pt 
\centering

\renewcommand{\arraystretch}{1.2}
\caption{\scriptsize Transfer and caching costs for different sizes of packed data items.}
\scriptsize
\begin{tabular}{|c|c|c|c|}
\hline
\textbf{\# Packed} & \textbf{Type} & \textbf{Transfer Cost} & \textbf{Caching Cost} \\ \hline

\multirow{2}{*}{1} 
& Unpacked & $\lambda$ & $\mu \cdot \Delta t$ \\
\cline{2-4}
& K-Packed & $\lambda$ & $\mu \cdot \Delta t$ \\ 
\hline

\multirow{2}{*}{2} 
& Unpacked & $2 \cdot \lambda$ & $2 \cdot \mu \cdot \Delta t$\\
\cline{2-4}
& K-Packed & $(1+\alpha) \cdot \lambda$ & $2 \cdot \mu \cdot \Delta t$  \\ \hline

\multirow{2}{*}{$|D_i|$} 
& Unpacked & $|D_i| \cdot \lambda$ & $|D_i| \cdot \mu \cdot \Delta t$ \\
\cline{2-4}
& K-Packed & $(1+(|D_i|-1)\cdot \alpha) \cdot \lambda$ & $|D_i| \cdot \mu \cdot \Delta t$ \\ \hline

\end{tabular}

\label{data_table}
\end{table}

 The process of packing and unpacking multiple data items can be time-consuming; hence, it is performed in advance to save time in online scenarios. The ESSs store single or multiple copies of the packed version of data items as shown in \autoref{fig:schema}. The CDN employs an optimized method for packing co-accessed data items, and ESSs utilize optimized methods for unpacking these data items. Due to these optimizations, the associated cost and latency are very less \cite{Wu23}. Conversely, data items that are not co-accessed and lack optimized packing methods, resulting in additional costs. Therefore, it is not beneficial to pack infrequently accessed data points; only co-accessed data points should be packed together.

\subsection{Problem Statement}

To satisfy a user data request, the CDN has two primary options: (a) Serve the request directly from the edge storage server (ESS) if the required data items are already cached locally. (b) Transfer the requested data items from another server (either another ESS or the cloud server) that holds the data, and subsequently cache the data locally.

The second approach involves replicating the data item(s), transferring the copy to the destination ESS, and caching it to ESS for future requests. This process incurs both transfer and caching costs. It is important to note that the fixed overheads associated with data replication and deletion are absorbed into the transfer or caching costs, allowing us to simplify the cost analysis without loss of generality.

The objective of the CDN is to minimize the total cost $C$ incurred in serving a sequence of user requests. Each request $r_i$ contributes to both caching cost ($C_P$) and transfer cost ($C_T$). The optimization problem can be formally stated as:

{\footnotesize
\begin{equation}
\min C = \sum_{r_i \in R} \left( C_T^{r_i} +  C_P^{r_i} \right)
\end{equation}}

The goal is to determine a serving and caching strategy that minimizes the overall cost while ensuring that all data requests are successfully fulfilled.

\section{Proposed Solutions}

\begin{figure}[tb!]
    \centering
    \includegraphics[scale=0.7]{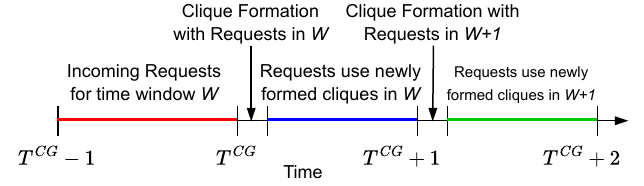}
    \caption{\footnotesize Timeline explanation for batching in online approach.}
    \label{fig:batch}
\end{figure}

Our proposed approach \textit{AKPC (Adaptive K-PackCache)} consists of three main modules, as outlined in \autoref{alg:packcache}. These include: (a) the \textit{Clique Generation Module} (\autoref{alg:construct_correlation_matrix}, \autoref{alg:alg_clique}, \autoref{alg:adjust_prev_cliques}), which identifies groups of frequently co-accessed data items (cliques) based on historical request patterns; (b) the \textit{Data Request Handling Module} (\autoref{alg:data_item_request_handling}), which processes incoming user requests by retrieving and serving packed or unpacked data items as appropriate; and (c) the \textit{Packed Copy Maintenance Module} (\autoref{alg:copy_expire}), which manages expiration and updates of cached packed data items.

The \textit{Clique Generation Module} is executed periodically every $T^{CG}$ time units and runs asynchronously in the background as shown in \autoref{fig:batch}. In contrast, the \textit{Data Request Handling Module} and \textit{Packed Copy Maintenance Module} operates in an event-driven manner, responding immediately to incoming user requests.

\subsection{Clique Generation Module}
\subsubsection{Construction of Normalized Correlation Matrix}

To identify item co-access patterns from recent user activity, we construct a normalized correlation matrix \( CRM_{Norm}(W) \) for the current time window \( W \), as described in \autoref{alg:construct_correlation_matrix}.

Each incoming request \( r_i = \langle D_i, s_i, t_i \rangle \in W \) contains a requested data items \( D_i \). Each data items belongs to $\mathcal{U}$ and $|\mathcal{U}|=n$. For every pair \( (i_1, i_2) \in \mathcal{U} \times \mathcal{U} \), the CDN increments the corresponding symmetric entries in a raw correlation matrix \( CRM(W) \) (\texttt{line 2-4}). After processing all requests, the matrix is normalized using min-max scaling to produce \( CRM_{Norm}(W) \in [0,1]^{n \times n} \), capturing relative co-access strength between item pairs (\texttt{line 5}).

To extract only the strongest associations, a binary matrix \( CRM^{bin}_{Norm}(W) \) is derived by thresholding the normalized values: entries above a given CRM threshold \( \theta \) are set to 1, and others to 0 (\texttt{line 6-9}). To reduce computational overhead, the CDN do not include all data items in the construction of \( CRM(W) \). Instead, it limits the scope to the top certain $\%$ of most frequently accessed data items in the current window. This results in a smaller, more focused matrix while still preserving high-impact co-utilization signals.

For example, suppose within window \( W \), the following two requests are observed: $r_1 = \langle \{d_1, d_2, d_3\}, s_1, t_1 \rangle, \: 
r_2 = \langle \{d_2, d_3\}, s_2, t_2 \rangle
$. Then co-access counts like \( CRM[d_2][d_3] \), \( CRM[d_3][d_2] \) will be incremented twice, while \( CRM[d_1][d_2] \), \( CRM[d_2][d_1] \) and \( CRM[d_1][d_3] \), \( CRM[d_3][d_1] \) will be incremented once. After normalization, if \( \theta = 0.4 \), and \( CRM_{Norm}[d_2][d_3] = 0.9 \), this edge will be retained in the binary matrix. 

\begin{figure}[tb!]
        \centering
        \includegraphics[scale=0.25]{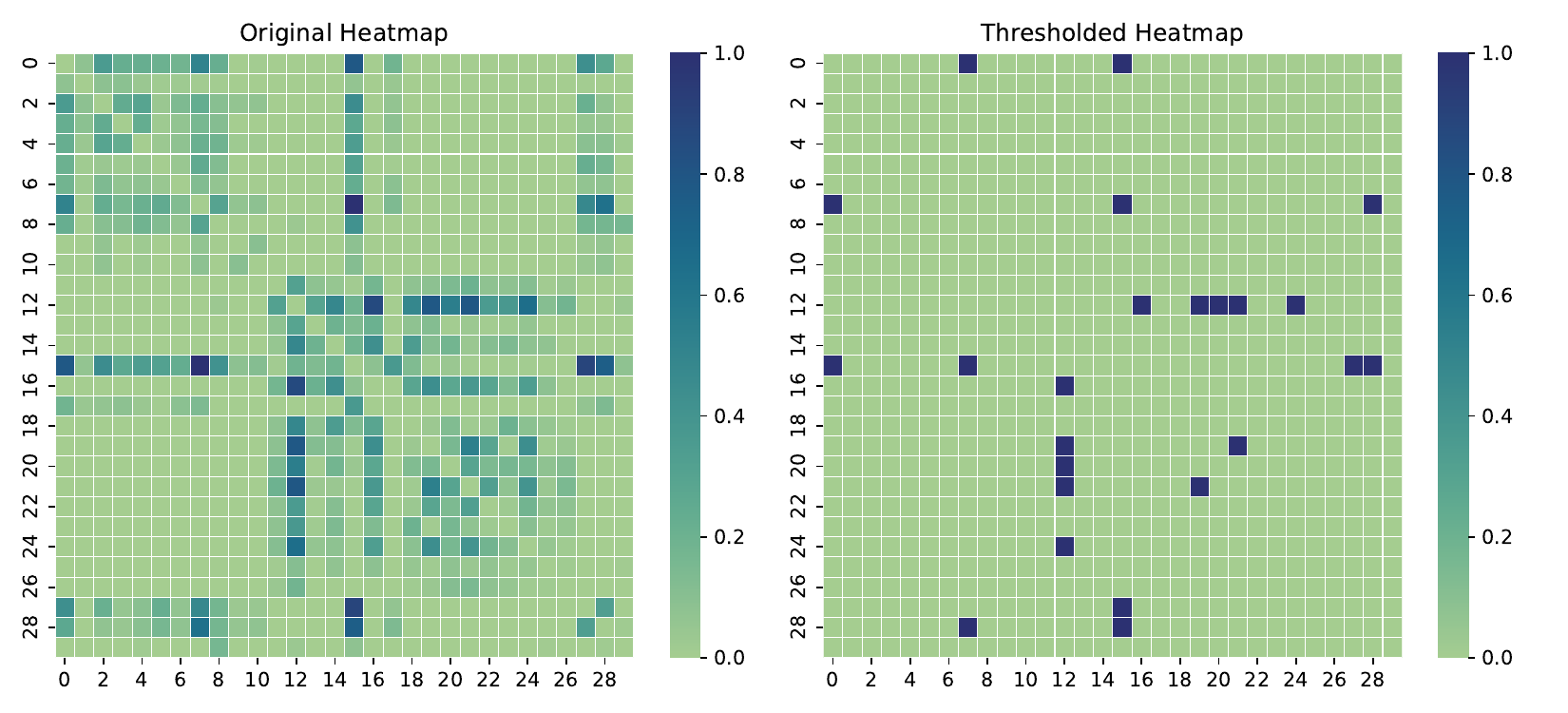}
        \caption{\footnotesize Heatmap depicting $CRM_{Norm}$ (on the left) and $CRM_{Norm}^{Bin}$ (on the right). Dataset used is top 30 \% used data item in Spotify dataset \cite{spotify}. It is autogenerated from code.}
        \label{fig:heatmap}
\end{figure}

A visual comparison of the normalized correlation matrix \( CRM_{\text{Norm}} \) and its binarized counterpart \( CRM_{\text{Norm}}^{\text{bin}} \), using the top 30\% most frequently accessed data items from the Spotify dataset~\cite{spotify}, is shown in \autoref{fig:heatmap}.

\begin{algorithm}[tb!]
\footnotesize
\caption{\footnotesize Adaptive K-PackCache Approach (\textit{AKPC})}
\label{alg:packcache}

\KwOut{Updated $C_P$, $C_T$}

\begin{algorithmic}[1]
\STATE{$C_P \gets 0$, $C_T\gets 0$, \texttt{Initialize }$C_P$ \texttt{(Caching Cost) and }$C_T$ \texttt{(Transfer Cost)} \texttt{to 0}}
%\FOR{every $T^{CG}$ interval}

\STATE{\textbf{Event 1:} \texttt{At every} $T^{CG}$ \texttt{time period}}
\STATE{Construct Normalized Matrix using \autoref{alg:construct_correlation_matrix}} ($W$, $r_i$, $n$)
\STATE{Generate Clique using \autoref{alg:alg_clique}} ($W$, $r_i$, $n$, $\omega$,$\theta$,$CRM^{bin}_{Norm}(W)$,$CRM^{bin}_{Norm}(W-1)$)
%\ENDFOR
\STATE{$G[c] \gets 1$, $E[c][j] \gets 1$ for a $s_j$ and $E[c][j] \gets 0$ for all other $s_j$, for every newly formed cliques $c \in$ $Clique(W)$}

%\IF{a request $r_i$ arrives at $s_j$}

\STATE{\textbf{Event 2:} \texttt{Request} $r_i$ \texttt{arrives at} $s_j$}
\STATE{Handle Data Request using \autoref{alg:data_item_request_handling}} ($r_i$,$Clique(W)$)
%\ENDIF
%\IF{a copy of clique $c$ expires at $s_j$ at $t_i$}
\STATE{\textbf{Event 3:} \texttt{Copy of clique} $c$ \texttt{expires at} $s_j$ \texttt{at} $t_i$}

    \STATE{Handling Clique Expiry using \autoref{alg:copy_expire} ($c$, $s_j$, $t_i$)}

%\ENDIF
\STATE {\textbf{return} $C_P$, $C_T$}

\end{algorithmic}
\end{algorithm}

\begin{algorithm}[tb!]
\footnotesize
    \caption{\footnotesize Construct Normalized Matrix}
    \label{alg:construct_correlation_matrix}
    \KwIn{$W \gets $Lists of all requests in time window $T^{CG}$, $r_{i}$:$\langle D_i, s_i, t_i \rangle\gets$ incoming requests, $\theta \gets$ CRM Threshold, $n \gets$ number of data points}
    \KwOut{$CRM_{Norm}(W)$: Normalized Correlation Matrix for requests in $W$}
    \begin{algorithmic}[1]
    \STATE{$CRM(W)$$[n]$$[n]$={0}}
    \FOR{{$r_i$} in $W$} 
        \FOR{{$i_1$, $i_2$} in $D_i$}
            \STATE{$CRM(W)$$[i_1]$$[i_2]$++, $CRM(W)$$[i_2]$$[i_1]$++}
        \ENDFOR
    \ENDFOR
    \STATE{$CRM_{Norm}(W)$ = min-max normalization on $CRM(W)$}
        \IF{$CRM_{Norm}(W)$ $>$ $\theta$}
        \STATE{$CRM_{Norm}^{bin}(W)$ = 1}
    \ELSE
        \STATE{$CRM_{Norm}^{bin}(W)$ = 0}
    \ENDIF
    \end{algorithmic}
\end{algorithm}

\begin{algorithm}[tb!]
\footnotesize
\caption{\footnotesize Disjoint Clique Construction with Reuse and Approximate Merging}
\label{alg:alg_clique}
\KwIn{
$W \gets$ Requests in current window, $r_i = \langle D_i, s_i, t_i \rangle$; $n \gets$ number of data points, $\omega \gets$ max (and target) clique size; $\theta \gets$ CRM threshold, $\gamma \gets$ Clique Approximation threshold, $CRM^{bin}_{Norm}(W-1) \gets$ previous binary CRM, $Clique(W-1) \gets$ cliques from previous window
}
\KwOut{$Clique(W)$: Disjoint set of cliques for window $W$}

\begin{algorithmic}[1]

\STATE $Clique(W) \gets$ adjust cliques using \autoref{alg:adjust_prev_cliques} ($Clique(W-1)$, difference of edge between $CRM^{bin}_{Norm}(W)$ and $CRM^{bin}_{Norm}(W-1)$)

\FOR{each $c \in Clique(W)$ and $|c| > \omega$}
    \STATE Split $c$ into smaller sub-cliques of size $\leq \omega$ using weakest co-utilization edges from $CRM_{Norm}(W)$
\ENDFOR

\FOR{all $c_1$ and $c_2$ in $Clique(W)$}
        \STATE $U \gets c_1 \cup c_2$
        \IF{$|U| == \omega$}
            \STATE $|E_U| \gets$ number of edges in $U$ from $CRM^{bin}_{Norm}(W)$
            \STATE $|E_{\max}| \gets \binom{\omega}{2}$ \texttt{\# edges for} $U$ \texttt{to be a clique}
            \IF{$|E_U| / |E_{\max}| \geq \gamma$}
                \STATE Update $Clique(W)$ by removing $c_1$, $c_2$ and adding $U$ (\textit{i.e.} newly generated combined clique)
            \ENDIF
        \ENDIF
\ENDFOR
\STATE \textbf{return} $Clique(W)$
\end{algorithmic}
\end{algorithm}

\begin{algorithm}[tb!]
\footnotesize
\caption{\footnotesize Adjust Previous Cliques}
\label{alg:adjust_prev_cliques}
\KwIn{
$Cliques(W-1) \gets$ Set of cliques from previous window ($W-1$), $\Delta E \gets$ Set of changed edges between $CRM^{bin}_{Norm}(W)$ and $CRM^{bin}_{Norm}(W-1)$
}
\KwOut{$Clique(W)\gets$ Updated list of cliques}

\begin{algorithmic}[1]
\STATE $Clique(W) \gets Clique(W-1)$

\FOR{each edge $(u, v)$ in $\Delta E$}
    \IF{edge $(u, v)$ was removed}
        \FOR{each clique $c$ in $Clique(W)$}
            \IF{$u, v \in c$}
                \STATE Remove $c$ from $Cliques(W)$
                \STATE Update $Cliques(W)$ by removing clique $c$ and add two newly formed cliques generated from removing edge $(u, v)$
            \ENDIF
        \ENDFOR
    \ELSIF{edge $(u, v)$ was added}
        \STATE Update $Cliques(W)$ if any new cliques are formed
    \ENDIF
\ENDFOR

\STATE \textbf{return} $Clique(W)$
\end{algorithmic}
\end{algorithm}

\begin{algorithm}[tb!]
\footnotesize
    \caption{\footnotesize Data Request Handling and Delivery}
    \label{alg:data_item_request_handling}
    \KwIn{Request $r_i$$\langle D_i, s_j, t_i \rangle$, $Clique(W) \gets$ cliques at window $W$}
    \KwOut{$Response(r_i)\gets$ set of cliques sent the the users}
    \begin{algorithmic}[1]

        \STATE{$Response (r_i) \gets \emptyset$ \texttt{Initialize set} $Response(r_i)$ \texttt{to Null}}
        \FOR{each $d \in D_i$}
            \STATE{Find the clique $c \in Clique(W)$ such that $d \in c$}
            \STATE{$Response \gets Response \cup c$}
            \STATE{$C_P\gets C_P+ |D_i|\cdot \mu \cdot((t_i+\Delta t)-E[c][j])$}
            \STATE{$E[c][j] \gets t_i +  \Delta t$ }
           
            \IF{$c$ is not present in $s_j$}
            \STATE{$G[c] \gets G[c] + 1$}
           
             \IF{$|c| == 1$}
           \STATE{$C_T\gets C_T + \lambda$}
            \ELSE 
            \STATE{$C_T\gets C_T + \alpha \cdot \mu \cdot |c|$}
            \ENDIF
            \ENDIF
        \ENDFOR
        \STATE{\textbf{return} all packed data items to users in $Response(r_i)$}
    
    \end{algorithmic}
\end{algorithm}

\subsubsection{Disjoint Clique Construction with Reuse and Approximate Merging}

In order to efficiently cluster correlated data items into co-accessed groups, we construct disjoint cliques using the binary correlation adjacency matrix \( CRM^{bin}_{Norm}(W) \) for the current request window \( W \). The procedure is detailed in \autoref{alg:alg_clique}.

The approach uses the normalized matrix \( CRM_{Norm}(W) \) and its binarized form \( CRM^{bin}_{Norm}(W) \) using \autoref{alg:construct_correlation_matrix} and incrementally updates cliques from the previous window \( Clique(W{-}1) \) using the procedure in \autoref{alg:adjust_prev_cliques}, which applies edge-level differences between the current and previous matrices.
\begin{itemize}
    \item \textit{Clique Splitting} \texttt{(line 2-3):} If a clique \( c \in Clique(W) \) exceeds the predefined size constraint \( \omega \), it is splited into smaller sub-cliques using the weakest co-utilization links based on \( CRM_{Norm}(W) \). For instance, suppose a clique \( c = \{d_1, d_2, d_3, d_4, d_5, d_6, d_7, d_8\} \) of size 8 (and \( \omega = 5 \)) exists, and the clique is broken into \( d_1,d_2,d_3,d_4 \) and \( \{d_5, d_6, d_7, d_8\} \) based on pairwise utilization data on $CRM_{Norm}(W)$.

\item \textit{Approximate Clique Merging} \texttt{(line 4--10):}  
To enhance packing compactness, we introduce an approximate merging strategy that reconstructs larger cliques of target size \( \omega \) by combining smaller cliques \( c_1 \) and \( c_2 \) under two specific conditions: (a) the union \( U = c_1 \cup c_2 \) must satisfy \( |U| = \omega \), and (b) the internal edge density of \( U \) must exceed a predefined approximation threshold \( \gamma \).  

Let \( |E_{\max}| = \frac{\omega(\omega - 1)}{2} \) be the maximum possible number of edges for a complete clique of size \( \omega \), and let \( |E_U| \) denote the actual number of edges in the subgraph induced by \( U \). If the ratio \( \frac{|E_U|}{|E_{\max}|} \geq \gamma \), then \( U \) is accepted as an approximate clique.  

This relaxed criterion enables near-cliques—those that are highly but not fully connected—to be treated as valid cliques, thereby improving packing efficiency. It reduces clique fragmentation and increases the chance of grouping co-accessed items, without significantly compromising on their correlation strength.

\end{itemize}

\subsubsection{Adjusting Previous Cliques}

This module is shown in \autoref{alg:adjust_prev_cliques}, incrementally updates cliques by examining edge changes \( \Delta E \) between the current ($CRM^{bin}_{Norm}(W)$) and previous binary correlation matrices ($CRM^{bin}_{Norm}(W-1)$) .

If an edge is removed, and both nodes of that edge exist in a clique \( c \), the clique is invalidated. It is removed, and two new cliques are formed by splitting along that edge. If an edge is added, and it leads to the creation of a new valid clique, that clique is added to the updated set. This approach avoids recomputing all cliques from scratch and exploits the temporal similarity of user behavior between successive time windows.

\subsubsection{Time Complexity}

Let \( n \) be the number of data items, \( k \) be the number of cliques, and \( \omega \) be the max clique size.
\begin{itemize}
\item Constructing \( CRM_{Norm}(W) \) takes \( O(|W| \cdot d_{\text{avg}}^2) \), where \( |W| \) is the number of requests and \( d_{\text{avg}} \) is the average number of items per request.

\item The \textit{Adjust Previous Cliques} approach involves scanning the changed edges \( \Delta E \), with each change affecting at most \( O(k) \) cliques. So, it runs in \( O(|\Delta E| \cdot k) \).

\item Splitting large cliques is based on utility-aware heuristics and takes \( O(k \cdot \omega^2) \), since each clique of size up to \( \omega \) is split using internal edge weights.

\item Approximate merging scans all \( \binom{k}{2} \) pairs of cliques, and for each pair, checks edge density in \( O(\omega^2) \) time. This yields a worst-case complexity of \( O(k^2 \cdot \omega^2) \).
\end{itemize}

The time-case complexity of \textit{Clique Generation Module} is \( O(k^2 \cdot \omega^2 + |\Delta E| \cdot k + |W| \cdot d_{\text{avg}}^2) \). In practice, due to temporal locality and reuse, both \( k \) and \( |\Delta E| \) are small, making the method efficient and suitable for online operation.

\subsection{Data Item Request Handling and Delivery}

This module is responsible for serving incoming user requests by selecting and delivering the appropriate packed cliques, based on the current set of disjoint cliques \( Clique(W) \) formed in the current time window. The detailed procedure is presented in \autoref{alg:data_item_request_handling}.

Each request \( r_i = \langle D_i, s_j, t_i \rangle \) corresponds to a set of data items \( D_i \) requested at time \( t_i \) from ESS \( s_j \). For each data item \( d \in D_i \), the CDN locates the corresponding clique \( c \in Clique(W) \) such that \( d \in c \), and includes \( c \) in the response set, denoted as \( Response(r_i) \). To avoid redundant transmission, cliques are treated as sets and included only once.

The CDN maintains two auxiliary counters for each clique \( c \) to support efficient state management in the \textit{K-PackCache} framework: (a) \( G[c] \): the number of active cached copies of \( c \) across all ESSs. (b) \( E[c][j] \): the expiration timestamp of \( c \) on ESS \( s_j \).

Each time a clique remains cached, its caching cost \( C_P \) increases by \( \mu \), the per-item per-unit time cost. Additionally, its expiration time \( E[c][j] \) is extended by a fixed interval \( \Delta t \) upon each access at server \( s_j \). Specifically, if a request for the clique is received at time \( t_i \), the expiration is updated to \( E[c][j] = t_i + \Delta t \) (\texttt{line 6}). Correspondingly, the caching cost \( C_P \) is incremented to reflect the extension, adding the cost associated with the additional interval beyond the current \( E[c][j] \) (\texttt{line 5}).

The transfer cost depends on the clique size: (a) If \( |c| = 1 \), the full transfer cost \( \lambda \) is applied (\texttt{line 10}). (b) If \( |c| > 1 \), a discounted cost \( \alpha \cdot \mu \cdot |c| \) is applied to represent amortized packed transmission cost, where \( \alpha \in [0, 1] \) is the discount factor (\texttt{line 12}).
If the clique is already cached at server \( s_j \), no additional transfer cost is incurred.

%Consider a request \( r_i = \langle \{d_2, d_3\}, s_1, t_i \rangle \), and suppose the current clique set contains \( c = \{d_2, d_3, d_4\} \). The server will respond with clique \( c \). If \( c \) is not cached at server \( s_1 \), a packed transfer cost of $(1 + (3 - 1) \cdot \alpha) \cdot \lambda$ is applied. All selected cliques are then sent together in a single response.

\textit{Time Complexity:} For a request involving \( |D_i| \) data items, the approach performs: (a) \( O(|D_i|) \) lookups to identify cliques, (b) \( O(|D_i|) \) updates to counters \( G[c] \), \( E[c][j] \), and cost values, (b) Constant-time checks for caching status. Thus, the overall time complexity is \( O(|D_i|) \), i.e., linear in the number of requested data items.

\subsection{Packed Copy Expiration Handling}
\begin{algorithm}[tb!]
\footnotesize
\caption{\footnotesize Copy Expire}
\label{alg:copy_expire}
\KwIn{Expired packed copy $c$ on server $s_j$ at time $t_i$}
\KwOut{$G[c]$, $E[c][j]$}
\begin{algorithmic}[1]
\STATE $\Delta t \gets \rho \cdot \frac{\lambda}{\mu}$
\IF{$G[c] == 1$ and $c\in Clique(W)$}
    \STATE $E[c][j] \gets t_i + \Delta t$ \texttt{Extend to prevent data loss}
\ELSE
    \STATE Drop copy of $d_p$ from $s_j$
    \STATE $E[c][j] \gets 0$, $G[c] \gets G[c] - 1$
\ENDIF
\STATE{\textbf{return} updated $G[c]$, $E[c][j]$}
\end{algorithmic}
\end{algorithm}

To ensure at one copy of clique is present in at least one ESS, the expiration handling logic is described in \autoref{alg:copy_expire}. In \texttt{line 1}, the value of $\Delta t$ is set to $\frac{\lambda}{\mu}$, where $\rho$ is a constant, similar to \cite{Wu23}.

When a packed copy \( c \) expires at server \( s_j \) and \( c \) is the last remaining copy in the entire system (\textit{i.e.}, \( G[c] = 1 \) and $c$ is in $Clique(W)$), its expiration is extended by \( \Delta t \) to prevent data loss (\texttt{line 3}). Otherwise, the copy is removed from \( s_j \), the expiration tracker \( E[c][j] \) is reset to 0, and the global counter \( G[c] \) is decremented (\texttt{line 6}).

Suppose \( c = \{d_2, d_3\} \) is cached on servers \( s_1 \) and \( s_2 \). When it expires at \( s_1 \), and \( G[c] = 2 \), the system deletes the copy from \( s_1 \) and updates \( G[c] \gets 1 \). If \( G[c] \) were already 1, the copy would be retained and its lifetime extended.

\textit{Time Complexity: } The expiration check and update operations are all performed in constant time. So, the time complexity of this module is \( O(1) \).

\subsection{Competitive Analysis}

The design of the \textit{K-PackCache} algorithm (AKPC) leads to the following key observations:

\textbf{Observation 1.} \textit{If $G[c_k] > 1$, i.e., more than one alive copy of clique $c_k$ exists, then no copy remains cached for more than $\Delta t$ on any server $s_j$, where $j \in [1, m]$.}

\textbf{Observation 2.} \textit{The algorithm has no knowledge of future requests. So, in the worst case, it caches $c_k$ for $\Delta t$ after serving a request. If no data point in $c_k$ is requested again during this period, the optimal would avoid caching it.}

\textbf{Observation 3.} \textit{Data loss does not occur. At any point, there is always at least one valid copy of every clique available\text{-}either locally cached or fetched from another server.}

\textbf{Observation 4.} \textit{Whenever a data request is received, the system delivers not only the requested data items but also all other data items belonging to the same cliques as those requested items. This implies that even unrequested data points may be transmitted as part of a packed clique.}

Based on the earlier observations, we now establish the competitive ratio of the \textit{AKPC} algorithm. Let $\mathcal{S}$ denote the number of data items requested by a user that are not cached locally at the serving storage server.

\textbf{Theorem 1:} \textit{The K-PackCache algorithm (AKPC) is} $
\frac{2 + (\omega - 1) \cdot \alpha \cdot \mathcal{S}}{1 + (\mathcal{S} - 1) \cdot \alpha}$-\textit{competitive}.

\textbf{Proof:} Let $C^i_{\text{AKPC}}$ and $C^i_{\text{OPT}}$ of the \textit{AKPC} and the optimal algorithm, respectively, for serving request $r_i$.

\textit{Case 1: Single Data Item Request: }Let $r_i = \langle \{d_i\}, s_j, t_i \rangle$ be a request for a single data item $d_i$ at server $s_j$. Assume $\lambda = \mu \cdot \Delta t$ (i.e., $\rho = 1$).

\textit{Case 1.1: $d_i$ is not cached at $s_j$: } AKPC sends the full clique $c$ of size at most $\omega$ for every data item requested. The transfer cost is $(1 + (\omega - 1)\cdot \alpha)\cdot \lambda$ and caching cost is $\mu \cdot \Delta t = \lambda$, giving:
$C^i_{\text{AKPC}} = (2 + (\omega - 1)\cdot \alpha) \cdot \lambda$. The optimal cost is only the transfer cost: $C^i_{\text{OPT}} = \lambda$. So,

{\footnotesize
\begin{equation*}
\frac{C^i_{\text{AKPC}}}{C^i_{\text{OPT}}} = 2 + (\omega - 1)\cdot \alpha \leq \frac{2 + (\omega - 1)\cdot \alpha \cdot \mathcal{S}}{1 + (\mathcal{S} - 1)\cdot \alpha}, \text{ with } \mathcal{S} = 1\end{equation*}}

\textit{Case 1.2: $d_i$ is cached at $s_j$: } Both algorithms incur caching cost only which is $\mu \cdot \Delta t$. So, we conclude:

{\footnotesize
\begin{equation*}
\frac{C^i_{\text{AKPC}}}{C^i_{\text{OPT}}} = 1 \leq \frac{2 + (\omega - 1)\cdot \alpha \cdot \mathcal{S}}{1 + (\mathcal{S} - 1)\cdot \alpha}
\end{equation*}}

\textit{Case 2: Multi-Item Request: } Let $r_i = \langle D_i, s_j, t_i \rangle$ denote a request where $|D_i| > 1$. Let $\mathcal{S}$ be the number of data items in $D_i$ that are not cached locally at server $s_j$. For simplicity, we assume $\mathcal{S} \leq \omega$. If $\mathcal{S} > \omega$, the request can be logically split into multiple sub-requests arriving at the same time and server, each satisfying the size constraint.

In the worst-case scenario, each of the $\mathcal{S}$ uncached data items belongs to a distinct clique. Due to Observation 4, the \textit{AKPC} algorithm transfers the entire clique even when only a single item from that clique is requested. Consequently, $\mathcal{S}$ separate clique transfers are triggered. Each transferred clique may contain up to $\omega$ data items, and the transfer cost incorporates a discount factor $\alpha$ for the additional $\omega - 1$ items per clique. 

In contrast, the optimal algorithm, having future knowledge, can pack all the $\mathcal{S}$ requested data items into a single clique transfer, thereby incurring only one discounted cost proportional to the number of items actually requested.

\textit{Case 2.1: All $|D_i|$ items are missed: } For this case, we consider $\mathcal{S}=|D_i|$. \textit{AKPC} transfers each missed item's full clique (up to size $\omega$). So, the transfer cost is $\mathcal{S} \cdot (1 + (\omega - 1)\cdot \alpha)\cdot \lambda$ and caching cost is $\mathcal{S} \cdot \mu \cdot \Delta t = \mathcal{S} \cdot \lambda$. So the total cost for \textit{AKPC} is: $
C^i_{\text{AKPC}} = \mathcal{S} \cdot (2 + (\omega - 1)\cdot \alpha) \cdot \lambda$. The optimal cost for transferring the $\mathcal{S}$ items packed together is:$
C^i_{\text{OPT}} = (1 + (\mathcal{S} - 1)\cdot \alpha) \cdot \lambda
$. So we conclude,

{\footnotesize
\begin{equation*}
\frac{C^i_{\text{AKPC}}}{C^i_{\text{OPT}}} = \frac{2 + (\omega - 1)\cdot \alpha \cdot \mathcal{S}}{1 + (\mathcal{S} - 1)\cdot \alpha} \leq \frac{2 + (\omega - 1)\cdot \alpha \cdot \mathcal{S}}{1 + (\mathcal{S} - 1)\cdot \alpha}
\end{equation*}}

\textit{Case 2.2: All requested items are cached: } Like \textit{Case 1.2}, both algorithm incur caching cost only which is $|D_i| \cdot\mu \cdot \Delta t$. So we can conclude:

{\footnotesize 
\begin{equation*}
\frac{C^i_{\text{AKPC}}}{C^i_{\text{OPT}}} = 1 \leq \frac{2 + (\omega - 1)\cdot \alpha \cdot \mathcal{S}}{1 + (\mathcal{S} - 1)\cdot \alpha} 
\end{equation*}
}

\textit{Case 2.3: Partial Cache Miss ($\mathcal{S} < |D_i|$): } The transfer cost is $\mathcal{S} \cdot (1 + (\omega - 1)\cdot \alpha)\cdot \lambda$ and caching cost is $\mathcal{S} \cdot \mu \cdot \Delta t = \mathcal{S} \cdot \lambda$. So the total cost for \textit{AKPC} is: $
C^i_{\text{AKPC}} = \mathcal{S} \cdot (2 + (\omega - 1)\cdot \alpha) \cdot \lambda$. The optimal cost for transferring the $\mathcal{S}$ items packed together is:$
C^i_{\text{OPT}} = (1 + (\mathcal{S} - 1)\cdot \alpha) \cdot \lambda
$. So we conclude,

{\footnotesize
\begin{equation*}
\frac{C^i_{\text{AKPC}}}{C^i_{\text{OPT}}} = \frac{2 + (\omega - 1)\cdot \alpha \cdot \mathcal{S}}{1 + (\mathcal{S} - 1)\cdot \alpha}  \leq \frac{2 + (\omega - 1)\cdot \alpha \cdot \mathcal{S}}{1 + (\mathcal{S} - 1)\cdot \alpha}
\end{equation*}}

\textit{Conclusion: } For any request $r_i$, the competitive ratio satisfies:

{\footnotesize
\begin{equation*}
\frac{C^i_{\text{AKPC}}}{C^i_{\text{OPT}}} \leq \frac{2 + (\omega - 1)\cdot \alpha \cdot \mathcal{S}}{1 + (\mathcal{S} - 1)\cdot \alpha}
\end{equation*}}

\textbf{Theorem 2.} \textit{No deterministic online algorithm under the \textit{AKPC} model can achieve a better competitive ratio than:} $\frac{2 + (\omega - 1) \cdot \alpha \cdot \mathcal{S}}{1 + (\mathcal{S} - 1) \cdot \alpha}$.

\textbf{Proof: }We construct an adversarial request sequence in phases $l_1, l_2, \ldots, l_k$ at a fixed server $s_j$, where each request $r_i = \langle D_i, s_j, t_i \rangle$ contains $|D_i| = \mathcal{S}$ data items that are uncached at the time of the request. We assume worst-case conditions where: (a) Each data item in $D_i$ belongs to different clique. (b) Each clique has size exactly $\omega$. (c) No item is ever requested again (Observation 2).

Phase $l_1$: Initial Request: (a) Request $r_1$ for $\mathcal{S}$ new items at $s_j$, all uncached, (b) \textit{AKPC} transfers full cliques for each of the $\mathcal{S}$ items, (c) Each transfer costs $\lambda \cdot (1 + (\omega - 1) \cdot \alpha)$ and caching costs $\lambda$.

Hence, $C^{l_1}_{\text{AKPC}} = \mathcal{S} \cdot (2 + (\omega - 1) \cdot \alpha) \cdot \lambda$. The optimal algorithm transfers all $\mathcal{S}$ items together: $C^{l_1}_{\text{OPT}} = (1 + (\mathcal{S} - 1) \cdot \alpha) \cdot \lambda$. The Competitive ratio for $l_1$ is $\frac{C^{l_1}_{\text{AKPC}}}{C^{l_1}_{\text{OPT}}} = \frac{2 + (\omega - 1) \cdot \alpha \cdot \mathcal{S}}{1 + (\mathcal{S} - 1) \cdot \alpha}$

Phase $l_2, l_3, \dots, l_k$: Adaptive Continuation: Now (a) Adversary waits until $t > t_{l_{i-1}} + \Delta t$ to ensure expiration of previous caches (Observation 1), (b) New request for $\mathcal{S}$ new data items, all disjoint and uncached, (c) \textit{AKPC} repeats same cost per phase:$C^{l_i}_{\text{AKPC}} = \mathcal{S} \cdot (2 + (\omega - 1) \cdot \alpha) \cdot \lambda$, $C^{l_i}_{\text{OPT}} = (1 + (\mathcal{S} - 1) \cdot \alpha) \cdot \lambda$. (d) The ratio remains same across all iterations. Total Cost After $k$ Iterations:

{\footnotesize
\begin{gather*}
\sum_{i=1}^{k} C^{l_i}_{\text{AKPC}} = k \cdot \mathcal{S} \cdot \left(2 + (\omega - 1) \cdot \alpha \right) \cdot \lambda, \\
\sum_{i=1}^{k} C^{l_i}_{\text{OPT}}  = k \cdot \left(1 + (\mathcal{S} - 1) \cdot \alpha \right) \cdot \lambda, \\
\frac{\sum_{i=1}^{k} C^{l_i}_{\text{AKPC}}}{\sum_{i=1}^{k} C^{l_i}_{\text{OPT}}} = \frac{2 + (\omega - 1) \cdot \alpha \cdot \mathcal{S}}{1 + (\mathcal{S} - 1) \cdot \alpha}.
\end{gather*}
}

\textit{Conclusion: } We can conclude that the adversary enforces the worst-case pattern under AKPC’s clique-based transfer strategy. Therefore, the derived bound is tight for any deterministic online algorithm operating under the \textit{AKPC} assumptions.

\subsection{Real-world Implementation}

Our work extends prior research on caching. Unlike \textit{PackCache} \cite{Wang17}, which co-utilizes two data items, we consider multiple data item caching based on co-utilization patterns. It enables an effective packing mechanism that reduces transfer and caching cost.

In contrast to \textit{DP\_Greedy} \cite{Huang19}, our \textit{AKPC} algorithm operates entirely in an online setting without any assumptions about future request sequences, and we derive a formal competitive ratio with a provable lower bound—neither of which are addressed in their work.

We also generalize the model in \textit{PackCache} \cite{Wu23}, which restricts each server to a single request per time step. Our design supports multiple concurrent requests per server closer to real-world workloads. Our approach responds by transmitting entire cliques, even if only some items are requested. This leverages data correlation for improved efficiency. Furthermore, while \cite{Wu23} tackles a version selection problem suitable for latency reduction, it introduces additional complexity. In contrast, \textit{AKPC} avoids version matching and is designed for fast, low-overhead real-time execution.

Finally, \textit{AKPC} builds upon the \textit{PackCache} framework \cite{Wu23} by supporting adaptive clique-based packing (up to size $\omega$), concurrent requests, and offering a tighter. This makes our algorithm more practical and scalable for modern content delivery and streaming systems.

\section{Experimental Evaluation}

To validate the practical usefulness of our proposed method, we carried out extensive experiments using real-world data. We developed a complete implementation of our approach in Python, which efficiently simulates the caching environment and applies our adaptive $K$-packing algorithm. This allowed us to test different scenarios, tune key parameters, and compare the performance of our method against existing baselines in a controlled and repeatable manner. 

%The code and datasets for this work are available at our \href{https://anonymous.4open.science/r/K-PackCache-10CF}{GitHub repository} \footnote{https://anonymous.4open.science/r/K-PackCache-10CF}.

\subsection{Datasets and Experimental Setup}

We conducted experiments on two real-world datasets: the Netflix trace~\cite{netflix} and the Spotify trace~\cite{spotify}. User location information was synthesized using realistic distribution patterns derived from prior studies~\cite{netflix_1, netflix_2, Vardo23}. Each request is defined as the set of data IDs accessed from a particular location (i.e., server) at a specific time instance.
To focus on high-impact access patterns, we extracted the top 10\% most frequently accessed data items from each dataset to calculate the CRM matrix. This filtering significantly reduces computational overhead while preserving the dominant co-utilization structure.  The size of each of the traces is 1 million. Details of the hyperparameter settings used in our experiments are summarized in \autoref{Table_hyperparameter}.

\begin{table}[tb!]
\centering
% \noindent\makebox[\textwidth]{%
\caption{\label{Table_hyperparameter} \footnotesize Base values used for the experiments}
\scriptsize
\begin{tabular}{ | c | c | } 
  \hline
  Parameters & Base values  \\
  \hline
    Cost ratio ($\rho$), $\mu$, $\lambda$ & 1, 1, 1 \\
    Max. Clique Size ($\omega$) & 5 \\
    Max. Request Size ($d^{max}$) & 5 \\
    Batch size & 200 \\
  CRM Threshold ($\theta$) & 0.2\\
  Clique Approx. Threshold ($\gamma$) & .85 \\
 Discount Factor $\alpha$ & 0.8, simlar to \cite{Huang19,Wu23} \\
  No. of servers ($|\mathcal{S}|=m$) & 600 \\
   No of data points ($|\mathcal{U}|=n$) & 60 \\

  \hline
\end{tabular}
% }

\end{table}

\begin{figure}[tb!]
    \centering
    \includegraphics[scale=0.6]{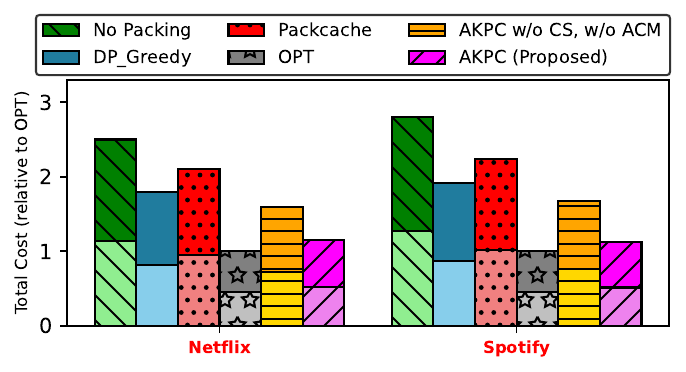}
    \caption{\footnotesize We compare the total cost of all state-of-the-art (SOTA) approaches against the proposed \textit{AKPC} variants. Each bar is split into two stacked components: the lower cylinder represents the transfer cost ($C_T$), while the upper segment denotes the caching cost ($C_P$). The total height of the bar corresponds to the overall cost.}
    \label{fig:cost}
\end{figure}

\subsection{Baseline Approaches}

We compare our proposed approach against four baselines. 
(a) \textit{No Packing} sends each data item individually without grouping, resulting in high cost due to lack of optimization. It is inspired from \textit{Wang et al.}\cite{Wang17}. (b) \textit{DP\_Greedy}, inspired by Huang \textit{et al.} \cite{Huang19}, is an offline approach that performs 2-item packing based on the full request trace. (c) \textit{PackCache} is an online version of 2-packing proposed by Wu \textit{et al.} \cite{Wu23}, which incrementally updates pairwise packing decisions based on incoming requests. (d) \textit{OPT} is the optimal strategy that achieves the minimum possible cost using complete future knowledge.

We compare these approaches against our proposed method, \textit{AKPC} (Adaptive K-PackCache), as well as a reduced variant, \textit{AKPC w/o CS, w/o ACM}, which disables clique splitting (\textit{CS}) and approximate clique merging (\textit{ACM}). All costs are reported relative to the optimal (\textit{OPT}), which is normalized to 1.

\subsection{Cost Evaluation Across Packing Scenarios and Baselines}
\textit{1. Cost Comparison with State-of-the-Art Packing Methods: } \autoref{fig:cost} presents the normalized total cost comparison across various packing strategies on the Netflix and Spotify datasets. The \textit{No Packing} strategy incurs the highest cost, as it fails to exploit any data co-access patterns, leading to redundant transfers and excessive caching. Both 2-packing methods — \textit{DP\_Greedy} (offline) and \textit{PackCache} (online) — improve upon this by pairing co-accessed items. However, they are limited by the fixed pairwise grouping, and cannot handle larger access correlations.

The proposed \textit{AKPC} (Adaptive K-PackCache) consistently achieves the lowest cost by forming variable-sized cliques of correlated items. It performs significantly better than prior methods, reducing cost by 63\% and 55\% compared to \textit{PackCache} for Netflix and Spotify respectively. It is only 15\% and 13\% higher than the offline OPT. Even the \textit{AKPC} variant without CS and ACM outperforms all existing baselines, highlighting the benefits of adaptive multi-item packing.

\begin{figure*}[htp]
  \centering
  \subfloat[\scriptsize Discount Factor ($\alpha$)]{\includegraphics[scale=0.51]{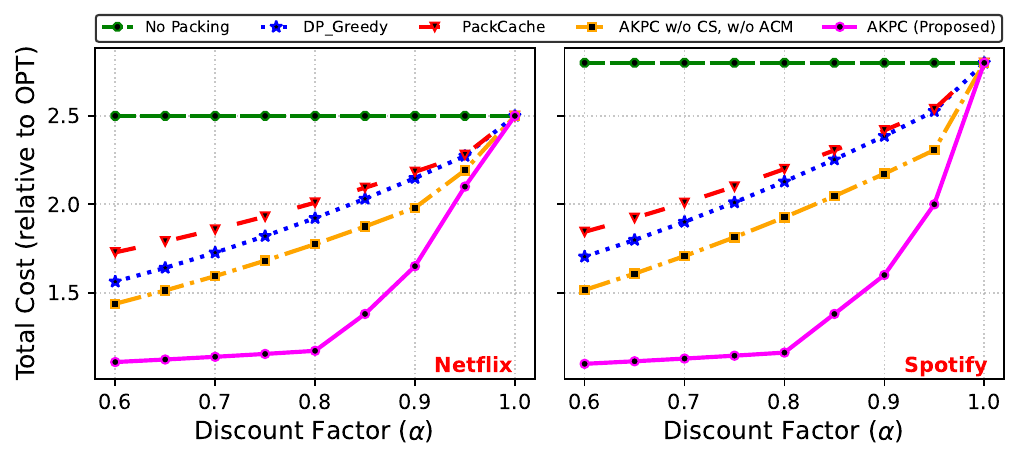}} 
  \hfill
  \subfloat[\scriptsize Cost Ratio ($\rho=\frac{\lambda}{\mu}$)]{\includegraphics[scale=0.51]{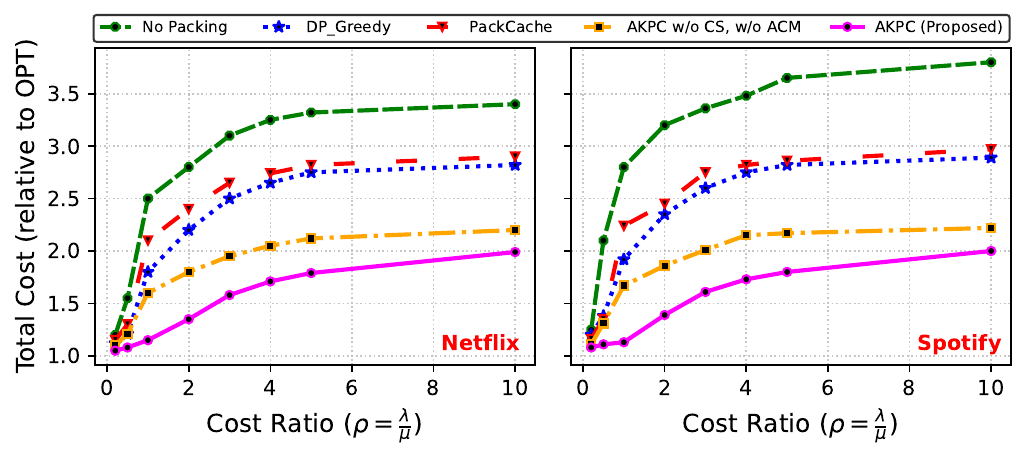}}
  \caption{\footnotesize Sensitivity analysis of relative total cost with respect to (a) Discount Factor ($\alpha$), which controls the packing benefit, and (b) Cost Ratio ($\rho = \lambda / \mu$).}
  \vspace{-0.5 cm}
  \label{fig:sen_analysis}
\end{figure*}

\textit{2. Impact of Discount Factor ($\alpha$): } \autoref{fig:sen_analysis} (a) illustrates how the relative cost increases with the discount factor $\alpha$ for both Netflix and Spotify datasets. The proposed \textit{AKPC} consistently achieves the lowest cost across all $\alpha$ values, especially when $\alpha \leq 0.85$, demonstrating its ability to leverage packing even when the cost benefit is marginal.
 
 As $\alpha$ rises from 0.6 to 1, the benefits of packing diminish, leading all approaches to converge toward the cost of the \textit{No Packing} baseline. This happens because lower values of $\alpha$ amplify the savings from jointly serving multiple co-accessed items. In contrast, higher $\alpha$ values reduce the marginal gain from packing, making all methods behave similarly.

\textit{3. Impact of Cost Ratio ($\gamma = \frac{\lambda}{\mu}$): } \autoref{fig:sen_analysis} (b) shows how the relative total cost varies with increasing cost ratio $\gamma$, which represents the ratio of transfer cost ($\lambda$) to caching cost ($\mu$), for both Netflix and Spotify datasets.

As $\gamma$ increases, the total cost rises for all approaches. This is expected since higher $\lambda$ values make data transfers more expensive. However, the proposed \textit{AKPC} method consistently incurs the lowest cost across all $\gamma$ values. This confirms that our approach is more effective when transfer costs dominate.

Notably, the benefit of \textit{AKPC} becomes more pronounced as $\gamma$ increases. At $\gamma=10$, \textit{AKPC} outperforms the best existing baseline by approximately 30\% on Netflix and 27\% on Spotify. This happens because our approach packs more data together, thereby reducing expensive transfers and maximizing caching efficiency.

\begin{figure*}[t]
  \centering
  \subfloat[\scriptsize CRM Threshold ($\theta$)]{\includegraphics[scale=0.51]{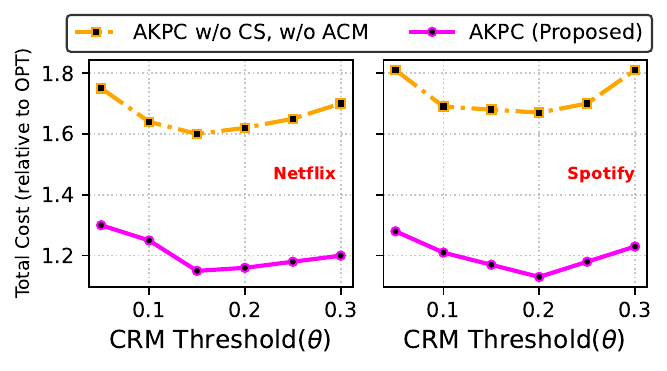}} 
  \hfill
  \subfloat[\scriptsize Clique Approx. Threshold ($\gamma$)]{\includegraphics[scale=0.51]{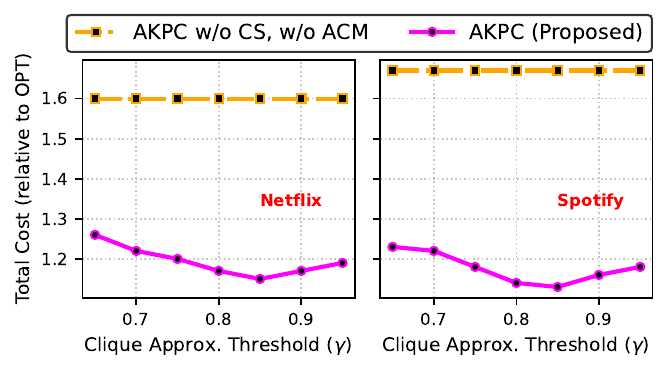}}\hfill
  \subfloat[\scriptsize Maximum Clique Size ($\omega$)]{\includegraphics[scale=0.51]{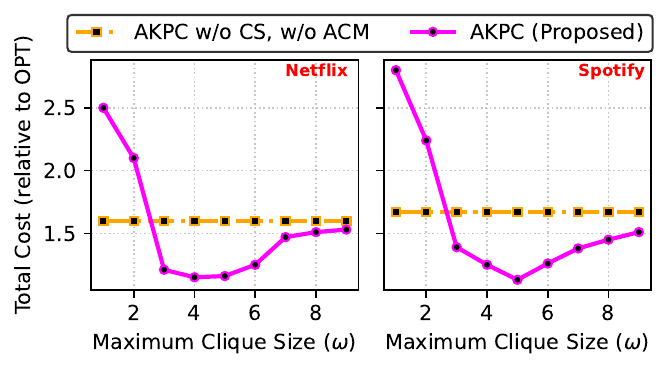}} 
  \caption{\footnotesize Hyperparameter analysis across different variations of the proposed approach. Variation of relative total cost with respect to (a) CRM threshold ($\theta$), (b) clique approximation threshold ($\gamma$), and (c) maximum clique size ($\omega$).}
  \label{fig:hyperparamter}
  \vspace{-0.5 cm}
\end{figure*}
\subsection{Impact of Hyperparameters on Cost Efficiency}

\textit{1. Impact of CRM Threshold ($\theta$): } The parameter $\theta$ determines the minimum edge weight required for forming connections in the correlation graph. A smaller value allows more edges, forming fewer but denser cliques, which may overlook some co-access opportunities. Conversely, a larger $\theta$ leads to the inclusion of weak correlations, resulting in the formation of unnecessary cliques that incur redundant caching costs. As shown in \autoref{fig:hyperparamter}(a), decreasing $\theta$ initially reduces the total cost by enabling effective packing. However, overly low thresholds begin to overfit weak patterns, increasing overall cost. The best values are $\theta=0.15$ for Netflix and $\theta=0.2$ for Spotify, balancing clique selectivity with packing efficiency. This behavior is consistent across both \textit{AKPC} and \textit{AKPC w/o CS, w/o ACM} variants.

\textit{2. Impact of Clique Approximation Ratio ($\gamma$): } This parameter $\gamma$ enables the inclusion of near-cliques during clique merging. A lower $\gamma$ permits looser grouping, increasing the number of approximate cliques, which may reduce transfer cost but lead to inefficient caching. A higher $\gamma$ enforces stricter merging, potentially missing useful co-access patterns. As shown in \autoref{fig:hyperparamter}(b), the cost initially decreases with increasing $\gamma$ due to better merges, then rises as valid groupings are lost. For both datasets, \textit{AKPC} achieves optimal performance at $\gamma = 0.85$, balancing merge precision and flexibility. For \textit{AKPC w/o CS, w/o ACM}, the cost remains constant, as this variant does not use clique approximation.

\textit{3. Impact of Maximum Clique Size ($\omega$): } The parameter $\omega$ limits the maximum number of items grouped into a single clique. Smaller values restrict packing opportunities, while larger values increase caching overhead. As shown in \autoref{fig:hyperparamter}(c), cost decreases initially with increasing $\omega$ due to better utilization of co-access patterns, but rises beyond $\omega = 5$ due to diminishing returns and added caching cost. The optimal value is $\omega = 5$ for both datasets in \textit{AKPC}. For \textit{AKPC w/o CS, w/o ACM}, this parameter has no effect, as it does not support multi-item cliques.

\subsection{Scalability Test}

\begin{figure*}[t]
  \centering
  \subfloat[\scriptsize Number of servers ($|\mathcal{S}|=m$)]{\includegraphics[scale=0.51]{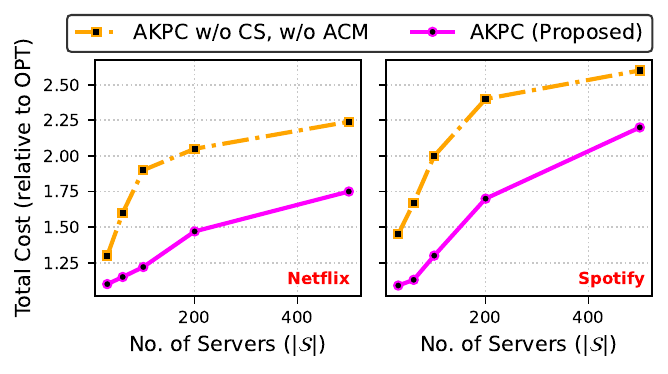}} \hfill
  \subfloat[\scriptsize Number of data points ($|\mathcal{U}|=n$)]{\includegraphics[scale=0.51]{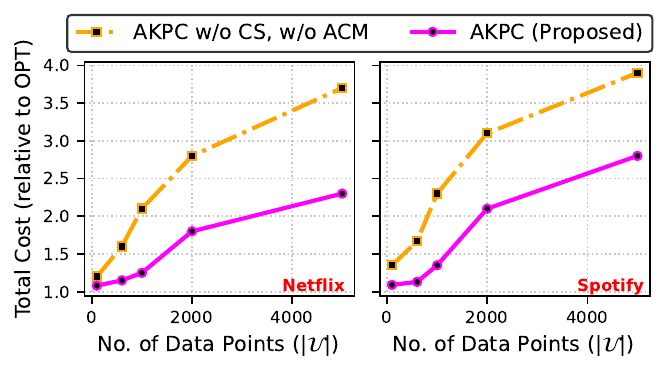}} \hfill
  \subfloat[\scriptsize Batch size]{\includegraphics[scale=0.51]{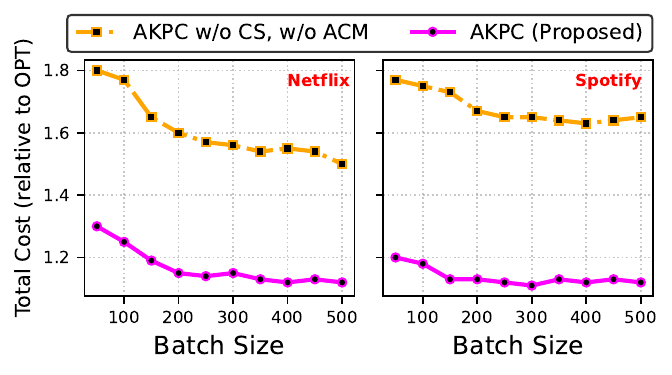}}
  \caption{\footnotesize Scalability analysis: Variation of relative total cost with (a) number of servers ($\mathcal{S}$), (b) number of data points ($\mathcal{U}$), and (c) number of requests in a batch.}
  \vspace{-0.5 cm}
  \label{fig:scalable}
\end{figure*}

\textit{1. Impact of Number of Servers: } As shown in \autoref{fig:scalable}(a), increasing the number of servers leads to a moderate increase in total cost. This is expected due to higher cumulative caching overhead. However, the increase is gradual — a 20$\times$ increase in servers results in only a 2$\times$ cost rise, indicating good scalability of the proposed method.

\textit{2. Impact of Number of Data Points: } From \autoref{fig:scalable}(b), we observe that as the number of data items grows, the total cost also increases. This is attributed to a higher chance of forming unnecessary or low-utility cliques. Still, the growth is controlled — a 60$\times$ increase in data points causes roughly a 4$\times$ increase in cost. Since clique formation focuses on frequently co-accessed items, the method maintains efficiency despite the growth.

\textit{3. Impact of Batch Size: } As seen in \autoref{fig:scalable}(c), increasing the batch size from 50 to 500 reduces the relative cost. Larger batches increase the likelihood of co-accessed items being grouped together, thus benefiting from the caching and transfer cost discounts. This trend confirms that the proposed \textit{AKPC} method effectively leverages batch-level co-utilization to reduce overall system cost.

\begin{figure}[t]
    \centering
    \includegraphics[scale = .45]{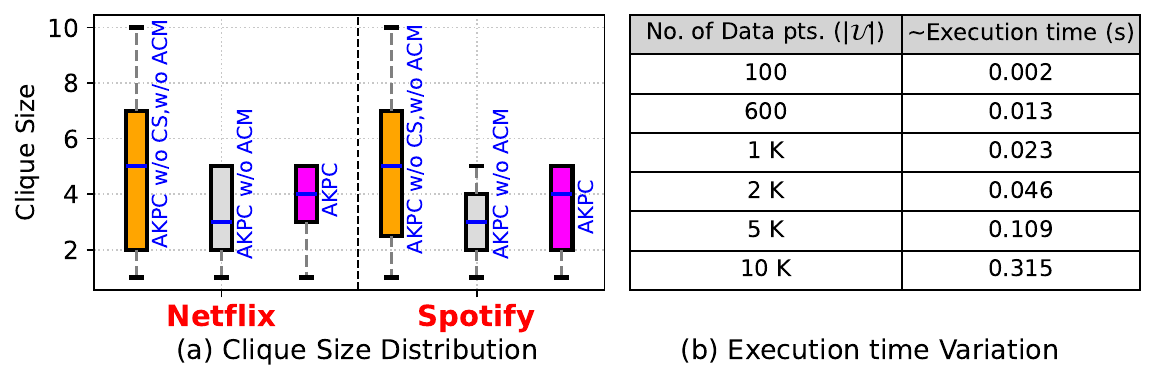}
    \caption{\footnotesize
        (a) Distribution of clique sizes across different variations of the proposed approach: 
        \textit{AKPC} without Clique Splitting and Approximate Clique Merging (denoted as \textit{AKPC w/o CS, w/o ACM}), 
        \textit{AKPC} with Clique Splitting only (denoted as \textit{AKPC w/o ACM}), 
        and \textit{AKPC} with both components (Proposed Method), and (b) Execution time (in seconds) of the proposed \textit{AKPC} method with varying sizes of data items. All experiments were conducted on a system with with Intel(R) Core(TM) i7-9700 CPU (8 cores), 16 GB RAM. 
    }
    \label{fig:akpc_box_table}
\end{figure}

\subsection{Analysing the Clique Size Distribution}

\autoref{fig:akpc_box_table} (a) shows the distribution of clique sizes across three variations of the proposed method: \textit{AKPC w/o CS, w/o ACM}, \textit{AKPC w/o ACM}, and the full \textit{AKPC} (Proposed). 

For \textit{AKPC w/o CS, w/o ACM}, the distribution is broad and centered, resembling a normal distribution for both Netflix and Spotify. This is expected, as the absence of both splitting and merging results in moderate-sized cliques, leading to a spread across the middle range. With \textit{AKPC w/o ACM}, we observe a skewed distribution with a clear upper bound, indicating that the maximum clique size is being more tightly controlled by the Clique Splitting mechanism. The distribution shifts slightly upward compared to the previous variant.

In the full \textit{AKPC} approach, the distribution shifts further upward. This is due to the Approximate Clique Merging (ACM) module, which introduces more higher-order cliques by relaxing the strict clique formation threshold. As a result, the average clique size increases, reflecting the aggressive merging strategy to optimize packing.

\subsection{Analysing the execution time of \textit{AKPC}}

We evaluate the runtime of our approach over varying data sizes, as summarized in Table~\ref{fig:akpc_box_table} (b). The measured runtime reflects only the computational overhead of our clique-based grouping algorithm, as the data access itself is virtually instantaneous. Importantly, this processing is not continuous but is triggered periodically in the background at fixed intervals to update the grouping structure in response to evolving access patterns.

The results indicate a stable and efficient runtime behavior: even for 10K data points, the algorithm completes within 0.32 seconds. This low-latency profile makes it well-suited for systems where updates need to happen incrementally without interrupting user-facing operations.

\section{Conclusion and Future Work}

In this work, we proposed \textit{AKPC}, a generalized $K$-packing framework for data caching that extends traditional 2-packing methods by leveraging clique-based strategies for grouping co-accessed items. The design incorporates approximate clique merging and splitting to improve packing efficiency while minimizing cache cost. Experiments on real-world Netflix and Spotify datasets show that \textit{AKPC} significantly reduces total cost and scales effectively with server count, data volume, and batch size—demonstrating the benefits of anticipatory, correlation-aware packing over baselines.

For future work, key directions include: (i) adaptive tuning of $K$ based on workload dynamics, (ii) modeling variable data sizes for heterogeneous storage settings, (iii) using online learning to adapt to shifting access patterns, and (iv) extending \textit{AKPC} to optimize energy use in domains such as IoT caching, edge clouds, and CDNs.

\footnotesize
\setstretch{0.85}
\bibliography{main}
\bibliographystyle{unsrt}

\vskip -2\baselineskip plus -1fil

\begin{IEEEbiography}[{\includegraphics[width=0.7in,height=0.7in]{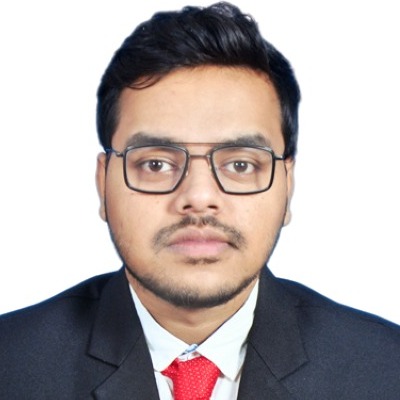}}]{Suvarthi Sarkar}
received his M.Sc. and M.Tech. degrees in Computer Science and Engineering from the University of Calcutta. He is currently pursuing a Ph.D. in Computer Science at the Indian Institute of Technology Guwahati, India. His research interests include cloud computing, real-time task scheduling, containerization, vehicular systems, and vehicular edge cloud computing.
\end{IEEEbiography}

\vskip -3\baselineskip plus -1fil

\begin{IEEEbiography}[{\includegraphics[width=0.7in,height=0.7in]{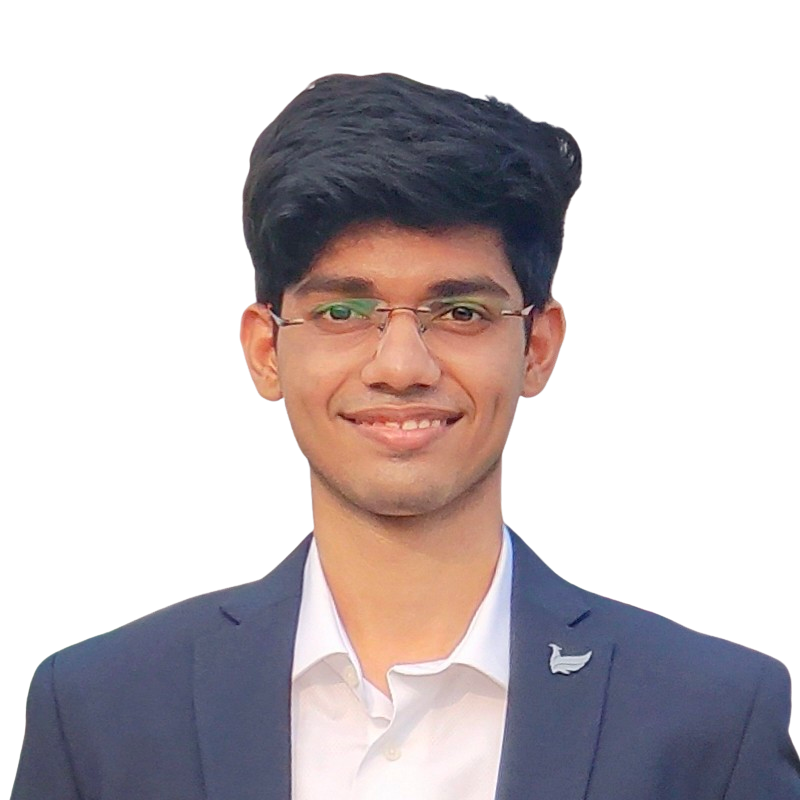}}]{Aadarshraj Sah}
received his B.Tech. major in Computer Science and Engineering and B.Tech. minor in Robotics \& Artificial Intelligence from the Indian Institute of Technology, Guwahati. His research interests include algorithms, optimization, and large-scale data-driven systems, and he has worked on deep learning for biomedical signal processing and rehabilitation applications. He is currently working as a Quant Researcher at QE Securities.
\end{IEEEbiography}

\vskip -3\baselineskip plus -1fil

\begin{IEEEbiography}[{\includegraphics[width=0.7in,height=0.7in]{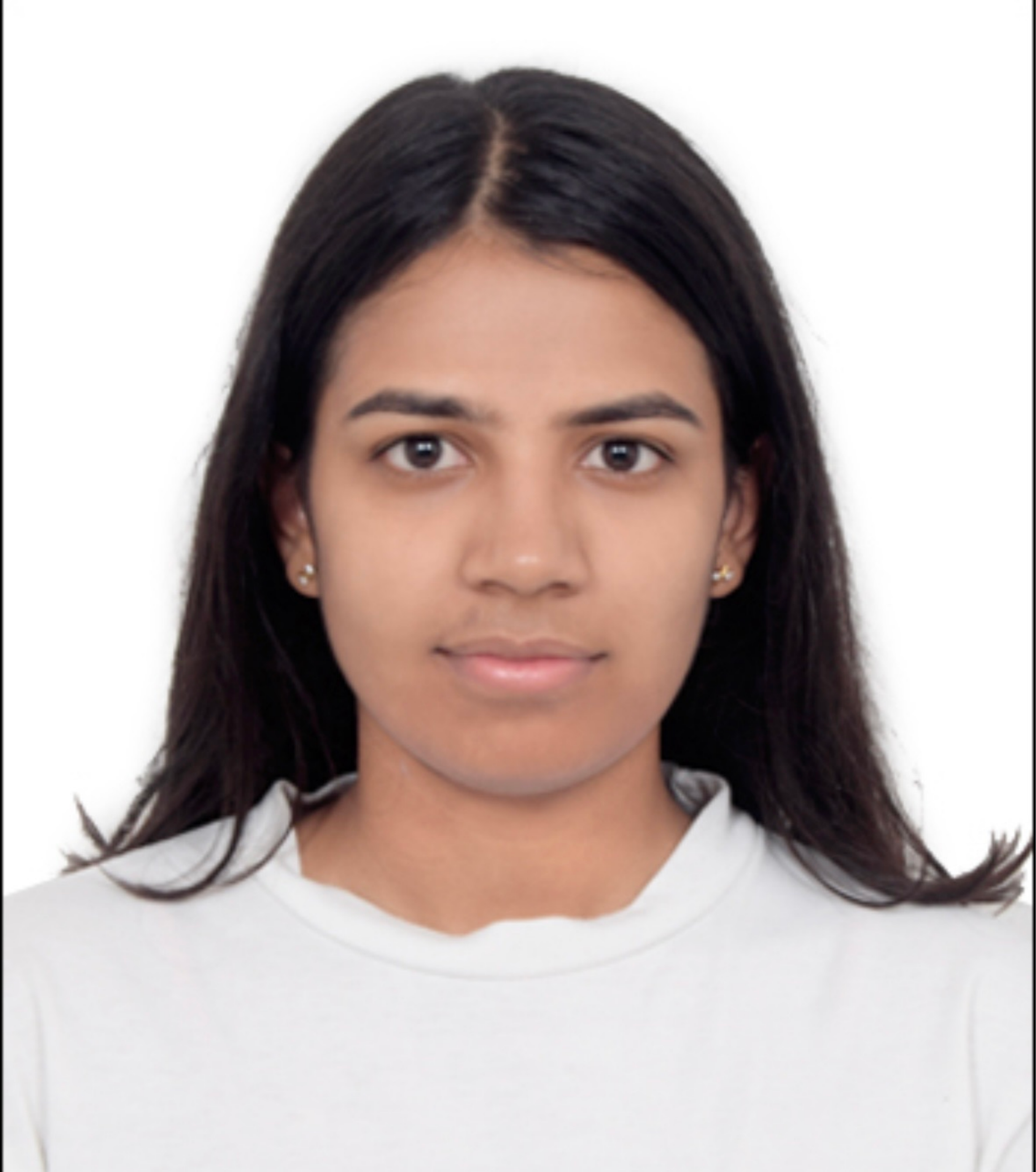}}]{Poddutoori Sweeya Reddy}
received her B.Tech. major in Computer Science and Engineering and B.Tech. minor in Robotics \& Artificial Intelligence from the Indian Institute of Technology, Guwahati. Her research interests focus on scalable computing systems and the application of large language models to automate complex engineering tasks. She is currently working as a Software Engineer at Google Cloud.
\end{IEEEbiography}

\vskip -3\baselineskip plus -1fil

\begin{IEEEbiography}[{\includegraphics[width=0.7in,height=0.7in]{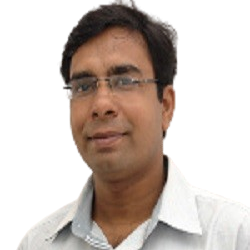}}]{Aryabartta Sahu}
received his Ph.D. in CSE from IIT Delhi. Since 2009, he has been serving as a faculty member in the Department of CSE at IIT Guwahati, India. He has authored over 35 research papers in reputed journals and conferences, including IEEE Transactions, Systems Journal, DATE, CCGrid, HPCC, etc. His research interests span advanced computer architecture, cloud systems, multicore parallel programming and compilation, embedded systems, as well as VLSI and FPGA design. He is actively involved in academic and professional communities and holds membership in ACM and senior membership in IEEE.
\end{IEEEbiography}

\end{document}